\documentclass[12pt]{article}

\usepackage[tbtags]{amsmath}
\usepackage{amssymb}
\usepackage{amsfonts}
\usepackage{euscript}

\newcommand{\mt}{\mathrm{tr}}
\newcommand{\nn}{\nonumber}

\begin{document}


\title{\large \bf  Local  invariants
for mixed qubit-qutrit states}

\author{\small V.~Gerdt\,${}^a$\,,\
        A.~Khvedelidze\,${}^{a,b}$\,, D. Mladenov \,${}^{c}$ and
        Yu.~Palii\,${}^{a,d}$
\\[0.7cm]
{${}^a$ } {\small \it Laboratory of Information Technologies,}\\
{\small \it Joint Institute for Nuclear Research, Dubna,
Russia} \\[0.3cm]
 ${}^b$ {\small
\it Department of Theoretical Physics,}\\
{\small \it A. Razmadze Mathematical Institute,} \\
{\small \it Tbilisi State University,
Tbilisi, Georgia} \\
[0.3cm]
 {\small {${}^c$
\it Department of Physics, Sofia State University, Sofia, Bulgaria
}}
\\[0.3cm]
 {\small {${}^d$
\it Institute of Applied Physics, Chisinau, Moldova}} }
\date{\empty}

\maketitle

\begin{abstract}
In the present paper few steps  are undertaken towards the
description of the \textit{\lq\lq qubit\--qutrit''} pair
\---  quantum bipartite system composed of two and three level
 subsystems. The computational difficulties with the construction of the
 ``\textit{local unitary polynomial  invariants}''  are discussed.
Calculations of the Molien functions and Poincar\'{e} series
for the qubit-qubit and  qubit-qutrit
local unitary
invariants  are  outlined and compared with the known results.
The requirement of positive semi-definiteness
of the density operator is formulated explicitly as a set of
inequalities in five Casimir invariants
of the algebra $\mathfrak{su}$(6).
\end{abstract}

\bigskip

\begin{center}
\end{center}

\newpage



\section{Introduction}


The present article  discusses several computational aspects
of a pure quantum effects in composite  systems playing an important role in
the modern theory of quantum computing and quantum information
\cite{Nielsen,Vedral}.

The cornerstone of these latest trends is an extraordinary quantum
phenomenon \--- the {\it``entanglement''} of quantum states.
Basically,  under the entanglement it is assumed an exposition of
diverse non-local correlations in a composite
multipartite  quantum system, which have no classical analogue. From
the mathematical standpoint of view characteristics of
entanglement can be understood  within the classical theory of
invariants (cf. \cite{Weyl,PopovVinberg}). The central object in these
studies is the ring of  G-invariant polynomials, called \textit{local invariants}, in elements of the
density matrix with the group G consisting  from the so-called
\textit{local unitary transformations} acting separately on every part of the
 multipartite composite system. The
program of description of this ring for multipartite
mixed states was outlined in \cite{LindenPopescu} and during the last decade
has been intensively developed. Over this time  many
interesting physical and pure mathematical results have been
obtained. Particularly,  for the simplest bipartite system of two
qubits,  the structure of the corresponding ring has been clarified
(see e.g. \cite{GrasslRottelerBeth,KingWelshJarvis,GPKPOMI09}).
However, comparative less is known for multipartite states, as well as for
bipartite mixed states, composed from arbitrary d-level subsystems, the
so-called qudits \cite{KusZyczkowski,BengtssonZyczkowski}. The reason is
first of all in a big computational difficulties we are faced. Indeed, even dealing
with 3-level subsystem,  qutrit,
the large number of independent elements of the density matrix
leads to the wide variety  of the local polynomial invariants and
makes non-effective the direct usage of the modern computer algebra software.

Below, attempting to construct the polynomial ring of
local invariants for qubit-qutrit pair, i.e. invariants against the action of SU(2)$\otimes$SU(3) group, we got added evidence of the complexity of the problem.
The known results~\cite{Djokovic} and our calculation of the Molien
function and Poincar\'{e} series show that
the number of local invariants grows up significantly compared with the two qubits case.
Nevertheless the derived  information is very useful
for the analysis of the polynomial ring of  local invariants.  As a preliminary result
we present here a set of linearly independent
local invariants up to the fourth order constructed via trace operation from the non-commutative monomials in  three elements of a special decomposition of qubit-qutrit density matrix. Using the subset of
the local invariants, consisting from the
Casimir invariants of the enveloping algebra
$\mathfrak{U}(\mathfrak{su}(6))\,,$
the positive semi-definiteness of density
matrix of qubit-qutrit pair is derived in the form of a system of algebraic
inequalities.

\section{The SU(n) Casimir invariants}

\label{sec:SUnCasimirs}

Here the basic statements on the unitary symmetry in quantum
mechanics and its role in the description of composite multipartite
states  is given.

{$\bullet$ \bf Density operator  and SU(n)-invariants $\bullet$ \ }
According to the conventional quantum theory, a complete information
about a generic n-dimensional system is accumulated in the self-adjoint
positive semi-definite density operator $\varrho\,$ with the unit
trace, $\varrho \in \mathfrak{P}_+$. For a closed quantum system,
this description is highly redundant, the equivalence relation
between  elements of $\mathfrak{P}_+\,,$  due to  the invariance of
observables  under the adjoint action of SU(n) group
\begin{equation} \label{eq:untran}
(\mathrm{Ad}\,g\,)\varrho  =  g\, \varrho\, g^{-1}\,, \qquad {g} \in
\mathrm{SU(\mathrm{n})}\,,
\end{equation}
guarantees that the physically relevant knowledge about quantum
states can be extracted from the orbit space $\mathfrak{P}_+\,
|$SU(n)~\footnote{The orbit space $\mathfrak{P}_+\,| $SU(n)\, of
SU(n) is defined  as the set of all SU(n)-orbits, endowed with the
quotient topology and differentiable structure,  the subset of
all the SU(n)-orbits with the same orbit-type forms a stratum of
$\mathfrak{P}_+\, |$SU(n)\,.}.
Relaxing for a moment condition of semi-definiteness, the density
operator $\varrho$  can be expressed via the Lie algebra $\mathfrak{su}(\mathrm{n})\,$  of  SU(n)
group~\cite{HioeEberly}:
\begin{equation}\label{eq:denalgebra}
    \varrho= \frac{1}{\mathrm{n}}\,\mathbb{I}_n +
   \tilde{\kappa}\,\imath\,\mathfrak{g}\,, \qquad
    \mathfrak{g} \in \mathfrak{su}(\mathrm{n})\,,
    \qquad \imath^2=-1\,.
\end{equation}
with some normalization factor $ \tilde{\kappa}\,.$ Therefore the  density
operator can be decomposed over $\mathrm{n}^2-1\,$ basis elements,
$e_i\,,$ of the Lie algebra
$\mathfrak{su}(\mathrm{n})\,$
\begin{equation}\label{eq:su(n)}
\mathfrak{g} = \sum_{i=1}^{\mathrm{n}^2-1}\,\xi_i\,e_i\,,
\end{equation}
and  any other operator $\mathcal{A}[\varrho]$\,, constructed from
the density operator $\rho$, admits a representation in  the graded
power series:
\begin{equation}\label{eq:obser}
\mathcal{A}(\boldsymbol{e}) = A^{(0)}\mathrm{I} + A^{(1)}_i\,e_i +
\frac{1}{2!}A^{(2)}_{ij}\,e_ie_j+
\frac{1}{3!}A^{(3)}_{ijk}\,e_ie_je_k + \dots \,.
\end{equation}
According to the Poincar\'e-Birkhoff-Witt theorem~\cite{Jacobson} the ordered monomials
\begin{equation}\label{eq:basis UEA}
   e_0=1\,,\  e_{i_1 i_2 \cdots i_k} =  e_{i_1} e_{i_2}\,\dots\, e_{i_k}\,,
   \qquad e_{i_1} < e_{i_2} <
   \dots <
   e_{i_k}\,,
\end{equation}
form  a linear basis of the universal enveloping algebra $\mathfrak{U}(\mathfrak{su}(\mathrm{n}))$
of $\mathfrak{su}(\mathrm{n})\,$.
Direct corollary of this theorem is that the symmetrized monomials of degree $d$ in (\ref{eq:obser})
span a linear spaces $\mathfrak{U}^d(\mathfrak{su}(\mathrm{n}))$ and
the universal enveloping algebra
\[
\mathfrak{U}(\mathfrak{su}(\mathrm{n}))=
\bigoplus_{d=0}^{\infty}\,\mathfrak{U}^d(\mathfrak{su}(\mathrm{n}))
\,.
\]
as a linear space is isomorphic to a polynomial algebra in commutative real variables
$\xi_i\,,\; i=1,\ldots,\mathrm{n}^2-1$.

Furthermore,  according to the well-known Gelfand's theorem
~\cite{Gelfand}, the description of center,
$\mathcal{Z}(\mathfrak{su}(\mathrm{n}))\,,$ of the enveloping
algebra $\mathfrak{U}(\mathfrak{su}(\mathrm{n}))$ reduces to the study of
invariants in  commutative symmetrized algebra
$S(\mathfrak{su(n)})\,$, which is isomorphic to the algebra of
invariant polynomials over $\mathfrak{su}(\mathrm{n})$. The elements
of center $\mathcal{Z}(\mathfrak{su}(\mathrm{n}))$ are in one to
one correspondence with the SU(n)-invariant polynomials in $\mathrm{n}^2-1\,$ real variables,
coordinates in $\mathfrak{su}(\mathrm{n})$. More precisely,
the element of $\mathfrak{U}(\mathfrak{su}(\mathrm{n}))$
\[
    \mathfrak{C}_r = \sum\,  \frac{1}{r!}\,
    c_{i_1 \cdots i_r}\sum_{\sigma \in \mathfrak{S}_r}\, e_{i_{\sigma(1)}} e_{i_{\sigma(2)}}\,
    \dots\, e_{i_{\sigma(r)}}\,,
\]
where $\mathfrak{S}_r$ is the group of permutation of $1,2, \dots r\,,$ belongs
to $\mathcal{Z}(\mathfrak{su}(\mathrm{n}))\,,$ if and only if $
c_{i_1 \cdots i_r}$  are coefficients of the polynomial in $\xi_1,
\xi_2, \dots , \xi_r$ variables
\[
    \phi(\xi_1, \xi_2, \dots , \xi_r) = \sum \, c_{i_1 \cdots i_r}\,
    \xi_{i_1} \xi_{i_2}\dots \xi_{i_r}\,,
\]
which is invariant under the adjoint action
\[
    \phi(\xi_1, \xi_2, \dots , \xi_r) =
    \phi((\mathrm{Ad}\ g)^T \xi_1, (\mathrm{Ad}\ g)^T \xi_2,
    \dots (\mathrm{Ad}\ g)^T\xi_r) \,,
\]
with $(\mathrm{Ad}\ g)^T$\-- the matrix of adjoint operator, $\mathrm{Ad}\ g$
, calculated in the basis $e_{i_1} e_{i_2}\,\dots\,
e_{i_r}$.

Therefore, from the algebraic standpoint,  the study of the orbit
space $\mathfrak{P}_+\, |$SU(n)\, as well as any characteristics of
quantum-mechanical observables, invariant under the unitary action
(\ref{eq:untran}),  reduces to the computation of the center
$\mathcal{Z}(\mathfrak{su}(\mathrm{n}))\,$ of $\mathfrak{U}(\mathfrak{su}(\mathrm{n}))$.

If the  elements $\mathfrak{C}_r\,$ belong to $\mathcal{Z}$ they are
termed as Casimir operators. The number of independent
homogeneous Casimir generators for SU$(\mathrm{n})$ group is equal to $\mathrm{rank} \
\mathfrak{su}(\mathrm{n})=\mathrm{n}-1$ .

It is well known, that the quadratic Casimir operator is unique up
to the constant factor and is expressible with the aid of  the
Cartan tensor:
\[
    C_{ij} = \mathrm{tr}((\mathrm{Ad}\ e_i)(\mathrm{Ad}\ e_j))\,,
\]
Therefore for algebra $\mathfrak{su}(\mathrm{n})$ the quadratic
Casimir operator reads
\[
    \mathfrak{C}_2 = \sum\, e_i e_i\,,
\]

The higher dimensional Casimirs  are expressed via the symmetric
structure constants $d_{ijk}\,$ of $\mathfrak{su}(\mathrm{n})\,$
algebra~\cite{Biedenharn}.
Because further, dealing with the qubit-qutrit system the Casimirs
of SU(6) will be used~\footnote{The tensorial
$\mathfrak{su}(2)\otimes\mathfrak{su}(3)$ product type basis for
$\mathfrak{su}(6)$  is given in Appendix \ref{sec:AppSU(6)}.
The formulas for the symmetric
structure constants $d_{ijk}\,$  and the antisymmetric
structure constants $f_{ijk}\,$ for $\mathfrak{su}(\mathrm{n})\,$  are  presented there  as well.},
the expressions for $\mathfrak{C}_i$ are given below:
\begin{eqnarray*}
  \mathfrak{C}_3 &=& \sum\, d_{i_1i_2i_3}
  \,e_{i_1}e_{i_2}e_{i_3}\,,\\
  \mathfrak{C}_4 &=&  \sum\, d_{ji_1i_2}d_{ji_3i_4}\,
  e_{i_1}e_{i_2}e_{i_3}e_{i_4}\,,\\
  \mathfrak{C}_5 &=&\sum\, d_{ii_1i_2}d_{iji_3}d_{ji_4i_5}\,
  e_{i_1}e_{i_2}e_{i_3}e_{i_4}e_{i_5}e_{i_6}\,,\\
  \mathfrak{C}_6 &=&\sum\, d_{ii_1i_2}d_{iji_3}d_{jki_4}d_{ki_5i_6}\,
e_{i_1}e_{i_2}e_{i_3}e_{i_4}e_{i_5}e_{i_6}\,.
\end{eqnarray*}

Now using these operators and decomposition (\ref{eq:su(n)}) based on the  isomorphism
between center
$\mathcal{Z}(\mathfrak{su}(\mathrm{n}))\,$ and SU(n)-invariant
polynomials, the following scalars,  referred hereafter as Casimir
invariants, can be written:
\begin{align}
    & \mathfrak{C}_2=(\mathrm{n}-1)\,\boldsymbol{\xi}\cdot\boldsymbol{\xi} \label{C2}\\
    & \mathfrak{C}_3=(\mathrm{n}-1)\,(\boldsymbol{\xi}\vee\boldsymbol{\xi}\,)\cdot\boldsymbol{\xi} \label{C3}\\
    & \mathfrak{C}_4=(\mathrm{n}-1)\,(\boldsymbol{\xi}\vee\boldsymbol{\xi}\,)\cdot(\boldsymbol{\xi}\vee\boldsymbol{\xi}\,) \label{C4}\\
    & \mathfrak{C}_5=(\mathrm{n}-1)\,\Big((\boldsymbol{\xi}\vee\boldsymbol{\xi}\,)
    \vee(\boldsymbol{\xi}\vee\boldsymbol{\xi}\,)\Big)\cdot\boldsymbol{\xi}
    \label{C5}\\
    & \mathfrak{C}_6=(\mathrm{n}-1)\,
    (\boldsymbol{\xi}\vee\boldsymbol{\xi}\vee\boldsymbol{\xi}\,)^2 \label{C6}
\end{align}
where
\[
    (\boldsymbol{U}\vee\boldsymbol{V}\,)_a:= \kappa\,d_{abc}U_bV_c\,,
\]
with normalization  constant
$\kappa:=\sqrt{{\mathrm{n}(\mathrm{n}-1)}/{2}}\,.$

Now, using these  scalars,   the positive semi-definiteness of density matrices for an arbitrary n-level
quantum system will be formulated.

{$\bullet$ \bf Positivity of density operators $\bullet$ \ }
To the best of our knowledge the first analysis of consequences of
the constraints on the density operator due to its  semi-positive
definiteness has been done in the sixtieth of the last century  studying
the production and decay of resonant states in strong interaction
processes \cite{Dalitz,Minnaert,DeenKabirKarl}. Nowadays, the
quantum computing and theory of quantum information reveals the
 new role of these constraints and recently they have been once more derived
 \cite{Kimura,ByrdKhaneja}~\footnote{In our recent publication~\cite{GPKPOMI09} the  positivity
 conditions for density operators
 has been analyzed in context of the consequences for
 integrity basis of
 SU(2)$\otimes$SU(2) polynomial invariants ring
 as well as for entanglement characteristics of mixed qubit states~\cite{GKPYadPhys10}.}.

To formulate the semi-definiteness let us choose the Bloch representation for a density operator~(\ref{eq:denalgebra})~\cite{HioeEberly}:
\begin{equation}\label{eq: Blochreprrho}
\varrho = \frac{1}{\mathrm{n}} \left(\,\mathbb{I}_{\mathrm{n}} +\omega\,\right)\,,
\qquad \omega=\kappa\,\boldsymbol{\xi} \cdot \boldsymbol{\lambda}\,,
\end{equation}
where $(\mathrm{n}^2-1)\,$-dimensional Bloch vector $\boldsymbol{\xi} \in \mathbb{R}^{\mathrm{n}^2-1}\,$
is contracted with elements $\lambda_i\,,\;i=1,\ldots,\mathrm{n}^2-1\, $ of the Hermitian basis  of $\mathfrak{su}(\mathrm{n})$ Lie algebra.
According to \cite{Minnaert}~\footnote{Note that P.Minnaert
attributed the same result to D.N.Williams.}  a necessary and
sufficient condition for the Hermitian matrix to be positive is
that the coefficients $S_k$ of its characteristic equation
\begin{equation}\label{eq:chareq}
    |\mathbb{I}\,x-\varrho| =
    x^\mathrm{n}-S_1x^{\mathrm{n}-1} + S_2x^{\mathrm{n}-2}- \ldots +(-1)^\mathrm{n} S_\mathrm{n} =0
\end{equation}
should be non-negative
\begin{equation}\label{eq:poscoeff}
\varrho \geq 0 \quad \Leftrightarrow \quad S_k \geq 0 \qquad k=1,
\ldots, \mathrm{n}.
\end{equation}
It is convenient to rewrite these inequalities in terms of
normalized coefficients $\bar{S}_k:=S_k/\mathrm{max}\{\, S_k
\}\,.$  Since  the maximal values of $S_k$ correspond to a
maximally degenerate roots; $x_1=x_2=\ldots=x_\mathrm{n}=1/\mathrm{n}\,$ of the
characteristic equation (\ref{eq:chareq}) one can express them
via the binomial coefficients
\[\mathrm{max}\{\, S_k \} = \frac{1}{\mathrm{n}^k}\,\binom{\mathrm{n}}{\mathrm{n}-k}
    \]
and thus
\begin{equation}\label{eq:IneqSyst}
    0 \leq \bar{S}_k \leq 1 \qquad k=2,\ldots,\mathrm{n}\,.
\end{equation}

Now we are ready to rewrite the constraints (\ref{eq:IneqSyst}) in
terms of the Casimir invariants (\ref{C2})-(\ref{C6}). This is
possible since, each of three sets, $\mathfrak{C}_k$, or $S_k\,$, or $t_k = \mathrm{tr}\,(\varrho^k)\,$,
$k=2,\ldots,n$ forms the basis of algebraically independent invariants
of SU($n$) group~(see e.g. \cite{Biedenharn}).
The expressions for  the coefficients $S_k$ in terms of $t_m$ are well-known, they are given by determinants:
\[
S_k =\frac{1}{k!}
\begin{vmatrix}
  t_1 & 1 & 0 & 0 & \cdots & 0 \\
  t_2 & t_1 & 2 & 0 & \cdots & 0 \\
  t_3 & t_2 & t_1 & 3 & \cdots & 0 \\
  \vdots & \vdots & \vdots & \vdots & \ddots & \vdots \\
  t_{k-1} & t_{k-2} & t_{k-3} & t_{k-4} & \ldots & k-1 \\
  t_k & t_{k-1} & t_{k-2} & t_{k-3} & \ldots & t_1 \\
\end{vmatrix}\, .
\]
Further, $t_m$ can be represented as
polynomials in Casimir invariants.
Based on the expressions for traces of symmetrized products
of Lie algebra basis elements (see Appendix \ref{sec:AppSU(6)}, cf. also
\cite{ByrdKhaneja}) we have:
\begin{align}
    \mt\,(\omega^2) = & \mathrm{n}\mathfrak{C}_2\,,   \nonumber \\ 
    \mt\,(\omega^3) = & \mathrm{n}\mathfrak{C}_3\,,  \nonumber \\ 
    \mt\,(\omega^4) = & \mathrm{n}\,(\mathfrak{C}_2^2+\mathfrak{C}_4) \,, \nonumber\\ 
    \mt\,(\omega^5) = & \mathrm{n}\,(2\,\mathfrak{C}_2\mathfrak{C}_3+\mathfrak{C}_5)\,,
    \nonumber\\ 
    \mt\,(\omega^6) = & \mathrm{n}\,(\mathfrak{C}_2^3+2\,\mathfrak{C}_2\mathfrak{C}_4
                        +\mathfrak{C}_3^2+\mathfrak{C}_6)\,. \nonumber 
\end{align}
Finally, imposing the following normalization for the Casimir
invariants,
\begin{equation*}\label{CasimirNorm}
    C_k=\frac{(k-1)!}{(\mathrm{n}-1)(\mathrm{n}-2)\ldots(\mathrm{n}-k+1)}\,\mathfrak{C}_k\,,
\end{equation*}
we arrive at a system of inequalities in $\mathrm{su}(6)$ Casimir
invariants, that defines the positive semi-definiteness of the
density matrix of  qubit-qutrit pair:
\begin{equation}\label{eq:IneqSystSU6-2}
    0\leq C_2 \leq 1\, ,
\end{equation}
\begin{equation}\label{eq:IneqSystSU6-3}
    0\leq 3\,C_2-C_3 \leq 1\, ,
\end{equation}
\begin{equation}\label{eq:IneqSystSU6-4}
    0\leq 6\,C_2-5\,C_2^2 -4\,C_3+C_4 \leq 1\, ,
\end{equation}
\begin{equation}\label{eq:IneqSystSU6-5}
    0\leq (1-5\,C_2)^2 -30\,C_2C_3+10\,C_3-5\,C_4+C_5 \leq 1\,
\end{equation}
\begin{equation}\label{eq:IneqSystSU6-6}
    \begin{split}
      0\leq & (1-5C_2)^3 -180C_2C_3+125C_2C_4+ \\
        & +20C_3(1+5C_3)-15C_4+6C_5-C_6
     \leq 1\, .
    \end{split}
\end{equation}

Since the positive semi-definiteness of   density matrices plays  an exceptional role   in the entanglement quantification problem it is reasonable to rewrite the system of inequalities (\ref{eq:IneqSystSU6-2})-(\ref{eq:IneqSystSU6-6}) in terms of the local $\mathrm{SU(2)}\otimes\mathrm{SU(3)}$ invariants.


\section{The local unitary invariants}

\label{sec:LocUI}

\setcounter{equation}{0}

\noindent{$\bullet$ \bf The local invariance of composite states
$\bullet$ \ }
When a quantum system is obtained by combining  of $r$-subsystems
with
$\mathrm{n}_1, \mathrm{n}_2, \ldots , \mathrm{n}_r\, $ levels each,  the non-local properties of
the composite system  are in correspondence with a certain  decomposition of
the unitary operations (\ref{eq:untran}).

In order to discuss this decomposition consider the subgroup
of unitary group formed by the \textit{local unitary transformations}
\begin{equation}\label{eq:LUT}
\mathrm{SU(\mathrm{n}_1)}\otimes\mathrm{SU(\mathrm{n}_2)}\otimes\cdots\otimes
\mathrm{SU(\mathrm{n}_r)}\,,
\end{equation}
acting independently on the density matrix of each subsystem
\begin{equation*} \label{eq:untran2}
    \varrho^{(\mathrm{n}_i)}  \quad \to \quad {\varrho^{(\mathrm{n}_i)}}^\prime =
    {g} \varrho^{(\mathrm{n}_i)}\, g^{-1}\,
    \qquad g\in  \mathrm{SU}(\mathrm{n}_i)\,, \quad i=1,2,\ldots, r\,.
\end{equation*}

Two states of composite system connected by the local unitary transformations
(\ref{eq:LUT}) have the same non-local properties. The latter  can
be changed only by the rest of the unitary actions,
\begin{equation*}\label{eq:nonLUT}
\frac{\mathrm{SU(n)}}{\mathrm{SU(n_1)}\otimes\mathrm{SU(n_2)}\otimes\cdots\otimes
\mathrm{SU(n_r)}}\,,
\qquad \mathrm{n}=\mathrm{n}_1 \mathrm{n}_2 \cdots \mathrm{n}_r\,,
\end{equation*}
generating the class of non-local transformations.

Having this notions,  we are in position to discuss the structure of the corresponding ring of
polynomial local invariants, i.e. polynomials in elements of the
density matrices, which are scalars under the adjoint local unitary
transformations. It is well known that for any reductive linear algebraic
group G and for any finite dimensional G-module V,
the ring $\mathcal{R}^{\mathrm{G}}$ has
the Cohen-Macaulay property~\cite{HochsterRoberts} and possesses the  Hironaka decomposition
\begin{equation*}\label{eq:CohenMacaulyMatr}
\mathcal{R}^{\mathrm{G}}=\bigoplus_{a=0}^{r} \,J_a\,\mathbb{C}[K_1, K_2, \ldots,
K_{s}]\,,
\end{equation*}
where $K_b ,  b=1,2, \dots , s$ are  primary, algebraically
independent polynomials and
$J_a, \ a=0, 1, 2,  \dots , r, J_0=1, $ are secondary, linearly
independent invariants respectively.
According to that the corresponding Molien function $M_{\mathrm{G}}(q)$ for
$\mathcal{R}^{\mathrm{G}}$~\cite{KingWelshJarvis} can be expressed as follows
\begin{equation*}\label{}
   M_{\mathrm{G}}(q)=
   \frac{\sum_{a=0}^r q^{\deg J_a}}{\prod_{b=1}^s(1-q^{\deg K_b})}\, .
\end{equation*}
In this form it provides us with a certain knowledge on the numbers of algebraically
independent polynomials as well as linearly independent invariants.

{$\bullet$ \bf Molien functions for
$\mathbb{C}[\mathfrak{P}_+^{(2\otimes2)}]$ and
$\mathbb{C}[\mathfrak{P}_+^{(2\otimes3)}]$ $\bullet$ \ }
Let us start with a remark concerning the adjoint action
(\ref{eq:untran}). Consider  a non-degenerate density
matrices. In this case  using the natural identification of the
elements of a linear space spanned by the Hermitian
$\mathrm{n}\times\mathrm{n}$ matrices with the space
$\mathbb{R}^{\mathrm{n}^2-1}$
\[
\varrho  \to \rho_{ij}
\]
one can instead of the adjoint action (\ref{eq:untran}) consider the
linear representation on $\mathbb{R}^{\mathrm{n}^2-1}$
\[\, V_A^\prime = L_{AB}V_{B}\,,
\qquad\qquad  L_{AB} \in
\mathrm{SU({n})}\otimes\overline{\mathrm{SU({n})}} \,, \] where the
line over expression means the complex conjugation.

After  this identification in order to get some insight on the
structure of the ring of polynomial invariants of linear action of
Lie group $\mathrm{G}$ on the linear $V\,$ space we can compute the Molien function
\begin{equation}\label{eq:Molienfunct}
    M(\mathbb{C}[V]^{\mathrm{G}},q)=\int_\mathrm{G}\, \frac{d\mu(g)}{\det(\mathbb{I}-q\pi(g))}, \qquad
|q|<1\,,
\end{equation}
where  ${d\mu(g)}$ is the Haar measure for Lie group $\mathrm{G}$ and $\pi(g)$ is the
corresponding representation on $V\,$. We start with the system of
two qubits.

\noindent\underline{\bf \ Two qubits\,.  \ }
In this case the local unitary group is
\begin{equation*}\label{}
    \mathrm{G}=\mathrm{SU}(2)\otimes\mathrm{SU}(2)\,.
\end{equation*}
As it is well known for any reductive linear group  the integration
in (\ref{eq:Molienfunct}) reduces to the integration  over the
maximal compact subgroup $\mathrm{K}$ of
$\mathrm{G}$~\cite{PopovVinberg}. In the present case this results
in  integration over the maximal torus
\begin{equation*}\label{}
   \pi(g)=\mathrm{diag}\,( 1,1,z,z^{-1})\otimes \mathrm{diag}\,(1,1,w,w^{-1})\,,
\end{equation*}
where $z,w$ \-- coordinates on one-dimensional tori. Therefore
computations  reduce to the following two-dimensional integral
\begin{equation*}\label{}
    M_{\mathrm{SU}(2)\otimes\mathrm{SU}(2)}(q)=\frac{1}{(2\pi i)^2}
    \int_{|z|=1}\int_{|w|=1}\,\frac{d\,\mu}{\Psi(z,w,q)}\,,
\end{equation*}
where
\[d\,\mu=(1-z)^2(1-w)^2\frac{dz}{z^2}\frac{dw}{w^2}\,,\]
\begin{equation*}\label{}
    \det(\mathbb{I}-q\pi(g))=(1-q)\, \Psi(z,w,q)
\end{equation*}
\begin{equation*}\label{}
    \begin{split}
       \Psi(z,w,q) =&
       (1-q)^3(1-qz)^2(1-qw)^2(1-qz^{-1})^2(1-qw^{-1})^2\\
      & (1-qzw)(1-qz^{-1}w)(1-qzw^{-1})(1-qz^{-1}w^{-1})\,.\nonumber
    \end{split}
\end{equation*}
After integration we get the Molien function~\cite{KingWelshJarvis}
\begin{equation*}\label{}
   M_{\mathrm{SU}(2)\otimes\mathrm{SU}(2)}(q)=
   \frac{1 + q^4 + q^5 + 3 q^6 + 2 q^7 + 2 q^8 + 3 q^9 + q^{10} + q^{11} + q^{15}}
        {(1 - q^2)^3 (1 - q^3)^2 (1 - q^4)^3 (1-q^6)}\,,
\end{equation*}
which is the palindromic one,
\[
 M_{\mathrm{SU(2)}\otimes\mathrm{SU(2)}}({1}/{q}) =
-q^{15}M_{\mathrm{SU(2)}\otimes\mathrm{SU(2)}}(q)\,,
\]
with the degree consistent  with
\[\dim\,\mathrm{SU(4)}=15\,.\]

\noindent\underline{\bf Qubit-Qutrit\,.\ }
 Now the local unitary group is $\mathrm{G}=\mathrm{SU}(2)\otimes \mathrm{SU}(3)$ and owing
to the symmetries of the integrand (\ref{eq:Molienfunct})
the non-trivial part of the integration is entirely  accumulated in the diagonal
components of the $\pi(g)$-representation of the form:
\[
   \pi(g)_{\mathrm{diag}}=\mathrm{diag}\,( 1,1,x,x^{-1})\otimes
   \mathrm{diag}\,(1,1,1,y,z,yz,y^{-1},z^{-1},(yz)^{-1})\,,
\]
where $x,y$ and $z$ are coordinates on the maximal torus.
Therefore, the computation of the Molien function
(\ref{eq:Molienfunct}) reduces to the evaluation of the multiple
contour integral in complex planes over unit circles~\footnote{The
multiple integral (\ref{eq:M23count}) has been calculated using the
consecutive application of Cauchy's residue theorem. Since the
integrand $f$ has poles of rather high orders, computer computations
of the residues has been performed using the command {\tt Residue}
built-in Mathematica that implements the standard limit formula for
high order poles (see {\tt
http://mathworld.wolfram.com/ComplexResidue.html}).}:
\begin{equation}\label{eq:M23count}
    M_{\mathrm{SU(2)}\otimes\mathrm{SU(3)}}(q) = \frac{1}{(2\pi i)^3}
    \int_{|x|=1}\int_{|y|=1}\int_{|z|=1}\,f(x,y,z,q)dx\,dy\,dz\,,
\end{equation}
where
\[
    \begin{split}
         f(x,y,z,q)& =\frac{1}{xyz}
            \frac{(1-x^{-1})(1-y^{-1})(1-z^{-1})(1-(yz)^{-1})}{\Psi(x,y,z,q)}\,,
    \end{split}
\]
\[
    \det(\mathbb{I}-q\pi(g))=(1-q)\, \Psi(x,y,z,q)\,,
\]
and $\Psi(x,y,z,q) = $
\begin{equation*}\label{}
    \begin{split}
      & (1-q)^5(1-qy)^2(1-qz)^2(1-qyz)^2
      (1-\frac{q}{y})^2
      (1-\frac{q}{z})^2(1-\frac{q}{yz})^2\\
      & (1-qx)^3(1-qxy)(1-qxz)(1-qxyz)
      (1-\frac{qx}{y})(1-\frac{qx}{z})
      (1-\frac{qx}{yz}) \\
      &(1-\frac{q}{x})^3(1-\frac{qy}{x})
      (1-\frac{qz}{x})(1-\frac{qyz}{x})
        (1-\frac{q}{xy})(1-\frac{q}{xz})
        (1-\frac{q}{xyz})\,.\\
    \end{split}
\end{equation*}

As a result, the Molien function can be represented in the rational
form (cf.~\cite{Djokovic}):
\begin{equation}\label{eq:MolinQubitQutrit}
 M_{\mathrm{SU(2)}\otimes\mathrm{SU(3)}}(q) =
\frac{N}{D}\,,
\end{equation}
 where
\[
\begin{split}
  N=
& 1 + 4\, q^4 + 9\, q^5 + 38\, q^6 + 69\, q^7 + 173\, q^8 + 347\, q^9 + 733\, q^{10} + 1403\, q^{11} \\
& + 2796\, q^{12} + 5091\, q^{13} + 9286\, q^{14} + 16058\, q^{15} + 27208\, q^{16} + 44250\, q^{17} \\
& + 70537\, q^{18} + 108430\, q^{19} + 163158\, q^{20} + 238264\, q^{21} + 339974\, q^{22} \\
& + 472130\, q^{23} + 641187\, q^{24} + 848615\, q^{25} + 1098643\, q^{26} + 1388741\, q^{27} \\
& + 1717327\, q^{28} + 2075836\, q^{29} + 2456389\, q^{30} + 2843020\, q^{31} + 3222408\, q^{32} \\
& + 3575226\, q^{33} + 3884797\, q^{34} + 4133599\, q^{35} + 4308636\, q^{36} + 4398377\, q^{37} \\
& + 4398377\, q^{38} + \ldots +38\, q^{69} + 9\, q^{70} + 4\, q^{71} + q^{75} \\
D= & (1 - q^2)^3 (1 - q^3)^4 (1 - q^4)^5 (1 - q^5)^4 (1 -q^6)^5 (1 -
q^7)^2 (1 - q^8)\,.
\end{split}
\]
This Molien function is the palindromic one
\[
 M_{\mathrm{SU(2)}\otimes\mathrm{SU(3)}}({1}/{q}) =
q^{35}M_{\mathrm{SU(2)}\otimes\mathrm{SU(3)}}(q)\,,
\]
as provided by
\[\dim\,\mathrm{SU(6)}=35\,.\]

This form of the Molien function serves as source of information on
the polynomial ring of SU(2)$\otimes$ SU(3) invariants.
Particularly, one can endeavour to identify the structure of
algebraically independent local unitary scalars. According to
(\ref{eq:MolinQubitQutrit}) there are 24 independent scalars in
agreement with simple count of
\(\mathrm{dim}\left[\mathrm{SU(6)}/\mathrm{SU(2)}\otimes
\mathrm{SU(3)}\right]=35-11=24\, \). The set of these 24 polynomial
invariants may be composed from three invariants of degree 2, four
of degree 3, five of degree 4, four of degree 5, five of degree 6,
two of degree 7 and one of the degree 8.

Note that the Poincar\'e series of
$M_{\mathrm{SU(2)}\otimes\mathrm{SU(3)}}(q) $
\[
 M_{\mathrm{SU(2)}\otimes\mathrm{SU(3)}}(q)=
 \sum_{d=0}^{\infty}\,
 \mathrm{dim}\left(\mathcal{P}_{d}^{\mathrm{SU(2)}\otimes\mathrm{SU(3)}}\right)\, q^d
 \,,
\]
determines the number of homogeneous polynomial invariants of degree
$d$. According to the calculations of (\ref{eq:M23count}) the few
terms of the Taylor expansion  over $q$ are
\begin{eqnarray}\label{eq:Poncare23}
 M_{\mathrm{SU(2)}\otimes\mathrm{SU(3)}}(q)&=&
 1+3\,{q}^{2}+4\,{q}^{3}+15\,{q}^{4}+25\,{q}^{5}+90\,{q}^{6}+170\,{q}^
{7} +489\,{q}^{8}\nonumber\\
&+&1059\,{q}^{9}+2600\,{q}^{10} + 5641\,{q}^{11}+12872\,{q}^{12}
+27099\,{q}^{13}\nonumber\\
&+& 57990\,{q}^{14}+118254\,{q}^{15} +240187\,{q}^{16}+ O \left(
{q}^{ 17} \right)\,.
\end{eqnarray}

Now, having in mind the input from the structure of the Molien
function (\ref{eq:MolinQubitQutrit}), we attempt to construct
the local SU(2)$\otimes$SU(3) unitary invariants.

\noindent{$\bullet$ \bf Constructing SU(2)$\otimes$SU(3) invariants
$\bullet$ \ }
Let us introduce the decomposition for density matrices well
adapted to the case of composite qubit-qutrit system. The space
$\mathfrak{su(\mathrm{6})}$ in (\ref{eq:denalgebra}) for n=6
admits decomposition in the direct sum of three real spaces
\begin{equation}\label{eq:3spaceDec}
\mathfrak{su(\mathrm{6})} = \bigoplus_{a=1}^3 V_a =
\mathfrak{su(\mathrm{2})}\otimes\mathrm{I}_3 +
\mathrm{I}_2\otimes\mathfrak{su(\mathrm{2})}
 +
\mathfrak{su(\mathrm{2})}\otimes
\mathfrak{su(\mathrm{3})}\,. \nonumber
\end{equation}
Using Pauli matrices $\sigma_i$ 
as the basis for $\mathfrak{su(\mathrm{2})}$ and
Gell-Mann matrices $\lambda_a$ 
as the basis for $\mathfrak{su(\mathrm{3})}$
(see Appendix \ref{sec:AppSU(6)})
the density matrix~(\ref{eq: Blochreprrho}) for qubit-qutrit system can be
written as~\cite{KusZyczkowski,BengtssonZyczkowski}:
\begin{equation}\label{eq:rhoabc}
\varrho = \frac{1}{6}\,\left[\mathbb{I}_6+ \omega \right]\,, \qquad
\omega =\alpha +\beta+ \gamma\,, \nonumber
\end{equation}
where
\begin{equation}\label{eq:abc}
\alpha: = \sum_{i=1}^3 a_i\, \sigma_i\otimes\mathbb{I}_3\,, \qquad
 \beta:=\sum_{a=1}^8 b_a\, \mathbb{I}_2\otimes \lambda_a\,, \qquad
 \gamma:= \sum_{i=1}^3 \sum_{a=1}^8
 c_{ia}\,\sigma_i\otimes\lambda_a\,. \nonumber
\end{equation}

Among the 35=3+8+24  real parameters  $(a_i\,, b_a, c_{ia})$  the
first two sets, $a_i\,$ and $b_a\,,$  correspond to the  Bloch
vectors of an individual qubit and qutrit respectively; the
evaluation of partial trace yields  the reduced matrices for
subsystems:
\[
\varrho^{(A)}:=\mathrm{tr}_B(\varrho)=\frac{1}{2}(\mathbb{I}_2
    +\vec{a}\cdot\vec{\sigma}),\quad
\varrho^{(B)}:=\mathrm{tr}_A(\varrho)=\frac{1}{3}(\mathbb{I}_3
    +\vec{b}\cdot\vec{\lambda})\,,
\]
while the  variables $c_{ia}$ are entries of the so-called
correlation matrix \mbox{$C=(c_{ia})$}.

Now using a trace operation described below we can construct a set of local SU(2)$\otimes$SU(3) scalars,
candidates for the elements of the integrity basis.

In analogy with the  generators (\ref{eq:basis UEA}) of the universal enveloping algebra
we consider a set $\mathcal{M}$ of non-commutative monomials
\begin{equation}\label{eq:monomial}
\mathcal{M}_{i_1\dots{i_d}}:=
X_{i_1}\cdot X_{i_2}\cdot\ldots\cdot X_{i_d}\;\in\;\mathcal{M} \,, \nonumber
\end{equation}
where each of $X_{i_k}\,, k=1,\dots, d\,,$ is one of $\alpha\,,\beta\,,$ or $\gamma\,$.
To each $\mathcal{M}_{i_1\dots{i_d}}$ we assign a multidegree $(s,t,q),\; s+t+q=d\,,$ where  $s,t$ and $q$ are degrees of $\alpha\,,\beta\,,$ and $\gamma\,$ respectively.
The trace operation on monomials $\mathcal{M}_{i_1\dots{i_d}}$ 
\begin{equation}\label{eq:polmn}
\mathrm{tr}:\quad \mathcal{M}_{i_1\dots{i_d}}\;\rightarrow\;
\mathcal{P}_{stq}(a_i\,, b_a, c_{ia}):=
\mathrm{tr}\left(\mathcal{M}_{i_1\dots{i_d}}\right)\;\in\; \mathcal{P}\,.
\nonumber
\end{equation}
defines the map
$\mathrm{tr}: \; \mathcal{M} \to  \mathcal{P}\;$ of $\mathcal{M}$ into the algebra $\mathcal{P}$ of homogeneous
polynomials in variables $(a_i\,, b_a, c_{ia})$. A generic term of the
polynomial $\mathcal{P}_{stq}(a_i\,, b_a, c_{ia})$ 
is a convolution of
vectors $a_i\,, b_a$ and matrix $c_{ia}$ with traces
\[
\mathrm{tr}\left(\sigma_{1}\sigma_{2}\cdots\sigma_{p} \otimes
\lambda_{1}\lambda_{2}\cdots\lambda_{r} \right)
=\mathrm{tr}\left(\sigma_{1}\sigma_{2}\cdots\sigma_{p}
\right)\mathrm{tr}\left(
\lambda_{1}\lambda_{2}\cdots\lambda_{r} \right)\,,
\]
where $p=s+q$ and $r=t+q$.

Now it is easy to verify  that the images of the trace map are
local unitary invariants. Indeed,
since under the transformation of the form $k_1\otimes k_2 \,,$
where $ k_1 \in \mathrm{SU(2)}\,,$ and $ k_2 \in \mathrm{SU(3)}\,,$
the matrices $\sigma$'s and $\lambda$'s in the basis elements of
$\mathfrak{su}(6)$ (see Appendix \ref{sec:AppSU(6)})
are transformed independently, in adjoint manner
\[
\sigma \to k_1 \sigma k^{-1}_1\,, \qquad  \lambda \to k_2 \lambda
k_2^{-1}\,,
\]
the polynomials  $\mathrm{tr}\left(\mathcal{M}\right)$ are
invariant against SU(2)$\otimes$SU(3) action.

Therefore the polynomials $\mathcal{P}_{stq}(a_i\,, b_a, c_{ia})$
are the reserve for the integrity basis of the ring
$\mathbb{C}[\,\mathfrak{P}\,]^{\mathrm{SU(2)\otimes SU(3)}}\,.$ Now,
in contrast to the case of SU(n) Casimir invariants  built up with
the help of  symmetric structure constants only, dealing with the
scalars against  the tensor product of groups, the invariants are
constructed in terms of the antisymmetric structure constants as well. For example,
\[ \mathrm{tr}(\gamma^3)=
c_{ia}c_{jb}c_{kc}\mathrm{tr}\left(\sigma_i\sigma_j\sigma_k\otimes\lambda_a
\lambda_b\lambda_c\right)=c_{ia}c_{jb}c_{kc}\mathrm{tr}
\left(\sigma_i\sigma_j\sigma_k\right)\mathrm{tr}\left(\lambda_a
\lambda_b\lambda_c\right)\,.
\]
This quantity being invariant under the SU(2)$\otimes$SU(3) action
is expressible via totaly antisymmetric tensor $\epsilon_{ijk}$ \--
structure constants of $\mathfrak{su}(2)$ algebra and $f_{abc}$ \--
structure constants of $\mathfrak{su}(3)$:
\[
\mathrm{tr}(\gamma^3)= -4\,\mathrm{\varepsilon}_{ijk}
f_{abc}\,c_{ia}c_{jb}c_{kc}\,.
\]

Choosing a basis for local invariants, several types of
algebraic dependence between the polynomials in
$\mathcal{P}$ have to be taken into
account. It is worth to consider two illustrative examples. Applying
the Hamilton-Cayley theorem  for elements $\alpha, \beta$ and
$\gamma$, considered as Hermitian $6\times6$ matrices, one can
determine the algebraic identities  for the  polynomials of the degree
$d>7\,.$  Less obvious example of
relations between polynomials is due to the identities between the
structure constants of the algebra.~\footnote{For the detailed analysis of the relations of
that type we refer to \cite{Macfarlane}.}
Let us consider two invariants, both 4-th
order in variables $c_{ia}$ of the correlation matrix $C\,,$ but one constructed using the invariant
symmetric structure constants $d$ while the second one with the
anti-symmetric structure constants $f\,$:
\begin{eqnarray}
    \mathfrak{I}^{004}(dd)&=&d_{abc}\,d_{c
    pq}\,(C^TC)_{ab}(C^TC)_{pq} \nonumber
    \,,\\
{\mathfrak{I}}^{004}(ff)&=&f_{apc}f_{cbq}(C^TC)_{ab}(C^TC)_{pq}\,. \nonumber
\end{eqnarray}
With the aid of identities (\ref{eq:ff_dd}) and
(\ref{eq:ddsu3Ogievetskii}) (Appendix \ref{sec:AppSU(6)}) for the
structure constants of $\mathfrak{su}(3)$ algebra, one can convinced
that
\begin{equation}
 \mathfrak{I}^{004}(dd)= \frac{2}{3}\,{\mathfrak{I}}^{004}(ff) -
 \frac{1}{3}\left[ \left(\mathrm{tr}(C^TC)\right)^2-
 2\mathrm{tr}(C^TCC^TC)\right]\,. \nonumber
\end{equation}

 According to the Poincar\'{e} series
(\ref{eq:Poncare23}) there are 15 homogeneous scalars of order 4, while there are $81=3^4$ monomials in three
noncommutative variables. But since the elements $\alpha$
and $\beta$ commute this number reduces. Taking into account this
commutativity as well as  the invariance of trace operation under
the cyclic permutations of products, one can find 18  valuable monomials:
\begin{eqnarray}\label{eq:valmon}
&&\alpha^4\,, \beta^4\,,\gamma^4\,, \alpha^3\beta\,,
\alpha\beta^3\,, \alpha^3\gamma\,, \alpha\gamma^3\,,
\beta^3\gamma\,, \beta\gamma^3\,, \nonumber
\\
&&\alpha^2\beta^2\,, \alpha^2\gamma^2\,, \alpha\gamma\alpha\gamma\,,
\beta^2\gamma^2\,, \beta\gamma\beta\gamma\,, \nonumber
\\
&& \alpha^2\beta\gamma\,, \alpha\beta^2\gamma\,,
\alpha\beta\gamma^2\,, \alpha\gamma\beta\gamma\,. \nonumber
\end{eqnarray}
Taking traces of these monomials one can convince that five of them
form the kernel of trace map:
\begin{equation*}
\mathrm{tr}(\alpha^3\beta)=\mathrm{tr}(\alpha\beta^3)=\mathrm{tr}(\alpha^3\gamma)
=\mathrm{tr}(\beta^3\gamma)=\mathrm{tr}(\alpha^2\beta\gamma)=0\,, \nonumber
\end{equation*}
and images of last two monomials 
coincide up to sign
\[
\mathrm{tr}\,(\alpha\beta\gamma^2)
=-\mathrm{tr}\,(\alpha\gamma\beta\gamma)\,.
\]

Therefore the following set of twelve  traces
\begin{eqnarray}\label{eq:12polyn1} \nonumber
&&\mathrm{tr}\,(\alpha^4)\,,\  \mathrm{tr}\,(\beta^4)\,,\
\mathrm{tr}\,(\alpha^2\beta^2)\,,\ \mathrm{tr}\,(\alpha^2\gamma^2)\,,\
\\
&& \label{eq:12polyn2} \nonumber
\mathrm{tr}\,(\gamma^4)\,, \ \mathrm{tr}\,(\alpha\gamma^3)\,,\
\mathrm{tr}\,(\beta\gamma^3)\,,
\mathrm{tr}\,(\alpha\gamma\alpha\gamma)\,,\
\\
&&
\mathrm{tr}\,(\beta^2\gamma^2)\,,\
\mathrm{tr}\,(\beta\gamma\beta\gamma)\,,
\mathrm{tr}\,(\alpha\beta^2\gamma)\,,\
\mathrm{tr}\,(\alpha\beta\gamma^2)\,,\label{eq:12polyn3} \nonumber
\end{eqnarray}
plus three 4-th order polynomials constructed as products of second
degrees polynomials
$\mathrm{tr}\,(\alpha^2)\mathrm{tr}\,(\beta^2)\,, \
\mathrm{tr}\,(\alpha^2)\mathrm{tr}\,(\gamma^2)\,,\
\mathrm{tr}\,(\beta^2)\mathrm{tr}\,(\gamma^2)\,, $ are 15
homogeneous invariant polynomials in accordance with the
Poincar\'{e} series (\ref{eq:Poncare23}).

How difficult is it to extract the independent scalars from this list?
It is easy to verify that some traces 
are expressed in terms polynomials of second order; e.g.,
$\mathrm{tr}\,(\alpha^2\beta^2) = \tfrac{1}{6}\mathrm{tr}\,(\alpha^2)
\mathrm{tr}\,(\beta^2)\,.$
Concerning the remaining monomials one can see that several  of them
have the same
multidegree.  Namely, the following ``trace'' polynomials
\begin{enumerate}
  \item $\mathrm{tr}\,(\alpha^2\gamma^2)=\frac{1}{6}
\mathrm{tr}\,(\alpha^2)\mathrm{tr}\,(\gamma^2) \quad \mbox{and} \quad
\mathrm{tr}\,(\alpha\gamma\alpha\gamma)\,,$
  \item $\mathrm{tr}\,(\beta^2) \mathrm{tr}\,(\gamma^2)\,,
  \quad \mathrm{tr}\,(\beta^2\gamma^2) \quad \mbox{and} \quad
\mathrm{tr}\,(\beta\gamma\beta\gamma)\,,$
\end{enumerate}
belong to the same space $\mathcal{P}_{202}$ and  $\mathcal{P}_{022}$
respectively. Being linearly independent monomials, they obey the  following relations
\begin{eqnarray}\label{}
    &&\mathrm{tr}\,(\alpha^2\gamma^2)+\mathrm{tr}\,(\alpha\gamma\alpha\gamma)  = 8
    \,a_{i_1}a_{i_2}c_{i_1j_1}c_{i_2j_1}\,, \nn \\
    &&\mathrm{tr}\,(\beta^2\gamma^2) -
    \frac{1}{6} \mathrm{tr}\,(\beta^2) \mathrm{tr}\,(\gamma^2)  =
    4\,d_{j_1j_2k}\,d_{k j_3j_4}\,b_{j_1}b_{j_2}c_{i_1j_3}c_{i_1j_4}\,, \nn \\
    &&\mathrm{tr}\,(\beta^2\gamma^2) + \mathrm{tr}\,(\beta\gamma\beta\gamma) =
    8(\tfrac{2}{3}b_{j_1}b_{j_2}c_{i_1j_1}c_{i_1j_2}+
    d_{j_1j_2k}\,d_{k j_3j_4}\,b_{j_1}b_{j_3}c_{i_1j_2}c_{i_1j_4})\,, \nn
\end{eqnarray}
where summation over all indices is assumed. This circumstance leaves
an open question how to build  the elements of integrity basis
with a certain multidegree using the ``trace'' polynomials.

We resume our analysis by the following list of linearly
independent SU(2)$\otimes$SU(3)  scalars
which are not products of low orders ones~\footnote{Note that 2-nd and 3-d order invariants were proposed in~\cite{Djokovic}.}:
\begin{itemize}
\item degree 2, three  invariants
\[\mathrm{tr}(\alpha^2),\;
\mathrm{tr}(\beta^2),\;
\mathrm{tr}(\gamma^2)\,,\]

\item  degree 3, four invariants
\[\mathrm{tr}(\beta^3),\;\mathrm{tr}(\gamma^3),\;
\mathrm{tr}(\alpha\beta\gamma),\;
\mathrm{tr}(\beta\gamma^2)\,,\]

\item  degree 4, eight invariants
\begin{eqnarray}
&&
\mathrm{tr}\,(\gamma^4)\,, \ \mathrm{tr}\,(\alpha\gamma^3)\,,\
\mathrm{tr}\,(\beta\gamma^3)\,,
\mathrm{tr}\,(\alpha\gamma\alpha\gamma)\,,\ \nn
\\
&&
\mathrm{tr}\,(\beta^2\gamma^2)\,,\
\mathrm{tr}\,(\beta\gamma\beta\gamma)\,,
\mathrm{tr}\,(\alpha\beta^2\gamma)\,,\
\mathrm{tr}\,(\alpha\beta\gamma^2)\,. \nn
\end{eqnarray}

\end{itemize}

{$\bullet$ \bf Casimir invariants decomposition  $\bullet$ \ }
The expansion of the SU(6) Casimir invariants up to the 4-th order (\ref{C2})-(\ref{C4})
over the above suggested  SU(2)$\otimes$SU(3) ``trace'' scalars reads:
\begin{align}
    6\,\mathfrak{C}_2&=\mt\,(\alpha^2)+\mt\,(\beta^2)+\mt\,(\gamma^2)\,,  \nn \\
6\,\mathfrak{C}_3&=
\mt\,(\beta^3)+\mt\,(\gamma^3)+3\,\mt\,(\beta\gamma^2)+6\,\mt\,(\alpha\beta\gamma) \,, \nn
\end{align}
\[
\begin{split}
   6\,\mathfrak{C}_4 & = \frac{1}{3}\Big[
        \mt(\alpha^2) \Big(2\,\mt(\beta^2) + \mt(\gamma^2)\Big)
          + \tfrac{1}{4}\,\mt (\beta^2)^2 -  \tfrac{1}{2}\,\mt (\gamma^2)^2
           - \mt(\beta^2) \mt(\gamma^2) \Big] \\
     & \quad +4\Big[ \mt(\alpha\gamma^3) + \mt(\beta\gamma^3) +
     \mt(\beta^2\gamma^2) + \mt(\alpha\beta\gamma^2) +
        3\mt(\alpha\beta^2\gamma) \Big] \\
     & \quad + 2\Big[ \mt(\alpha\gamma\alpha\gamma) + \mt(\beta\gamma\beta\gamma) \Big] +\mt(\gamma^4) \,.
\end{split}
\]

We conclude with the final remark on the
applicability of the derived results to the problem of classification of  mixed quantum states.
Using inequalities
(\ref{eq:IneqSystSU6-2})-(\ref{eq:IneqSystSU6-6})
and results from~\cite{GKPYadPhys10} the well-known Peres---Horodecki
criterion for the separability  of qubit-qutrit
mixed states can be reformulated  as a set of inequalities
in SU(2)$\otimes$SU(3) scalars.

\bigskip
\bigskip

{$\bullet$ \bf Acknowledgments $\bullet$ \ }
The work was supported in part by the RFBR  (grant No. 10-01-00200),
and by Ministry of Education and Science of the Russian Federation
(grant No. 3810.2010.2) and JINR-Bulgaria 2010 collaborative grant.
The research of A.~K. was supported by
the GNSF research grant GNSF/ST08/4-405.



\appendix


\section{Appendix: Formulas for the  $ \mathfrak{su}(6)$ algebra }

\label{sec:AppSU(6)} 

\setcounter{equation}{0}

{$\bullet$ \bf The tensorial basis  $\bullet$ \ }
For the $\mathfrak{su}(6)$ algebra we use the basis $\{\tau_A\}_{A=1, \ldots,  35}$
constructed from the tensor products of
the Pauli matrices $\sigma_i \in \mathfrak{su}(2)$:
\[
    \sigma_1 =\left(%
\begin{array}{cc}
  0 & 1 \\
  1 & 0 \\
\end{array}
\right)\quad
\sigma_2 =\left(%
\begin{array}{cc}
  0 & -i \\
  i & 0 \\
\end{array}
\right)\quad
\sigma_3 =\left(%
\begin{array}{cc}
  1 & 0 \\
  0 & -1 \\
\end{array}
\right)\,,
\]
and eight $\{\lambda_a\}_{a=1,\ldots,8} $ Gell-Mann matrices,
forming the $\mathfrak{su}(3)$ basis:
\[
\begin{array}{c}
\begin{array}{ccc}
 \lambda _{1}=\left(\begin{array}{ccc}
   0 & 1 & 0 \\
   1 & 0 & 0 \\
   0 & 0 & 0 \
 \end{array}\right)&
 \lambda _{2}=\left(\begin{array}{ccc}
   0 & -i & 0 \\
   i & 0 & 0 \\
   0 & 0 & 0 \
 \end{array}\right)&
  \lambda _{3}=\left(\begin{array}{ccc}
   1 & 0 & 0 \\
   0 & -1 & 0 \\
   0 & 0 & 0 \
 \end{array}\right)\\ & & \\
  \lambda _{4}=\left(\begin{array}{ccc}
   0 & 0 & 1 \\
   0 & 0 & 0 \\
   1 & 0 & 0 \
 \end{array}\right)&
 \lambda _{5}=\left(\begin{array}{ccc}
   0 & 0 & -i \\
   0 & 0 & 0 \\
   i & 0 & 0 \
 \end{array}\right)&
  \lambda _{6}=\left(\begin{array}{ccc}
   0 & 0 & 0 \\
   0 & 0 & 1 \\
   0 & 1 & 0 \
 \end{array}\right)\
 \end{array}
 \\ \\
\begin{array}{cc}
\lambda _{7}=\left(\begin{array}{ccc}
   0 & 0 & 0 \\
   0 & 0 & -i \\
   0 & i & 0 \
 \end{array}\right)&
  \lambda _{8}=\displaystyle{\frac{1}{\sqrt{3}}}\left(
  \begin{array}{ccc}
   1 & 0 & 0 \\
   0 & 1 & 0 \\
   0 & 0 & -2 \
\end{array}\right)
\end{array}\
\end{array}
\]

The elements $\tau_ A\,$  are enumerated as
\begin{eqnarray}
&&\tau_i =\frac{1}{\sqrt{3}}\,\sigma_i\otimes\mathbb{I}_3\,,\qquad
\tau_{3+a} =\frac{1}{\sqrt{2}}\,\mathbb{I}_2\otimes\lambda_a\,, \nonumber
\\
&&\tau_{11+a}=\frac{1}{\sqrt{2}}\,\sigma_1\otimes\lambda_a\,,\
\tau_{19+a}=\frac{1}{\sqrt{2}}\,\sigma_2\otimes\lambda_a\,, \
\tau_{27+a}=\frac{1}{\sqrt{2}}\,\sigma_3\otimes\lambda_a\,.\nonumber
\end{eqnarray}

{$\bullet$ \bf The algebraic structures $\bullet$ \ }
The product of basis elements reads
\[
\tau_A\tau_B=\frac{2}{n}\delta_{AB}\mathbb{I}+(d_{ABC}+\imath\,f_{ABC})\tau_C\,,
\]
The structure constants $d_{ABC}$ and  $f_{ABC}$ can be determined
via equations
\[
d_{ABC} = \frac{1}{4}\mbox{Tr}(\{\tau_A,\tau_B\}\tau_C) , \qquad
f_{ABC} = -\frac{\imath}{4}\mbox{Tr}([\tau_A,\tau_B]\tau_C),
\]
where apart  from the Lie algebra product, $[\ \,, \ ]\,,$ the
``anti-commutator'' of  elements, i.e., \(\{\tau_A,\tau_B\}=\tau_A
\tau_B+\tau_B \tau_A \) has been used.

{$\bullet$ \bf Identities for structure constants $\bullet$ \ }
For the SU(n)  group the  the structure constants obey the following
identities:
\[
f_{abc} f_{cpq} + f_{bpc} f_{caq} + f_{pac} f_{cbq} = 0\,,
\]
\[
d_{abc} f_{cpq} + d_{bpc} f_{caq} + d_{pac} f_{cbq} = 0\,,
\]
\[
f_{abc} f_{cpq} = d_{apc} d_{cbq} - d_{aqc} d_{cbp} + \frac{2}{\mathrm{n}}\,
( \delta_{ap} \delta_{bq} - \delta_{aq} \delta_{bp} )\,,
\]
\begin{eqnarray}\label{eq:ff_dd}
f_{abc} f_{cpq} + f_{aqc} f_{cpb} &=& 2 d_{apc} d_{cbq} - d_{abc}
d_{cpq} - d_{aqc} d_{cbp}\nonumber\\
& +& \frac{2}{\mathrm{n}}\,( 2 \delta_{ap} \delta_{bq} - \delta_{ab}
\delta_{pq} - \delta_{aq} \delta_{bp} )\,.
\end{eqnarray}

The SU(3) symmetric constants satisfy
~\cite{OgievetskyPolubarinov,Bukhvostov} an important identities

\begin{equation}\label{eq:ddsu3Ogievetskii}
d_{abc}d_{cpq}+d_{bpc}d_{caq}+d_{pac}d_{cbq}  = \frac{1}{3}
(\delta_{ab}\delta_{pq}+\delta_{ap}\delta_{bq}+\delta_{aq}\delta_{bp})\,.
\end{equation}

{$\bullet$ \bf The traces   $\bullet$ \ }
The traces of symmetrized products of $\mathfrak{su}(\mathrm{n})$
basis elements are
\begin{align}\label{}
    \mathrm{tr}\,(\tau_{\{a}\tau_{b\}}) & =2\,\delta_{ab}\,, \nonumber \\
    \mathrm{tr}\,(\tau_{\{a}\tau_{b}\tau_{c\}}) & =2\,d_{abc}\,, \nonumber \\
    \mathrm{tr}\,(\tau_{\{a}\tau_{b}\tau_{c}\tau_{d\}}) & =
    \frac{2^2}{\mathrm{n}}\,\delta_{ab}\delta_{cd}
                +2\,d_{abe}d_{ecd}\,, \nonumber \\
    \mathrm{tr}\,(\tau_{\{a}\tau_{b}\tau_{c}\tau_{d}\tau_{e\}}) & =
                  \frac{2^2}{\mathrm{n}}\,(d_{abc}\delta_{de}+\delta_{ab}d_{cde})
                  +2\,d_{abf}d_{fcg}d_{gde}\,, \nonumber \\
    \mathrm{tr}\,(\tau_{\{a}\tau_{b}\tau_{c}\tau_{d}\tau_{e}\tau_{f\}}) & =
              \frac{2^3}{\mathrm{n}^2}\,\delta_{ab}\delta_{cd}\delta_{ef}
            +\frac{2^2}{\mathrm{n}}\,(d_{abg}d_{gcd}\delta_{ef}+
            \delta_{ab}d_{cdg}d_{gef}) + \nonumber \\
       &  \qquad +\frac{2^2}{\mathrm{n}}\,d_{abc}d_{def}
        +2\,d_{abg}d_{gch}d_{hdv}d_{vef} \,. \nonumber
\end{align}



\begin{thebibliography}{99}
%
\bibitem{Nielsen}M.A.~Nielsen and I.L.~Chuang, \textit{Quantum Computation
and Quantum Information}, Camridge University Press, 2000.
%
\bibitem{Vedral}V.~Vedral, \textit{Introduction to Quantum Information Science},
 Oxford University Press, New York, 2006.
%
\bibitem{Weyl}H.~Weyl, \textit{The Classical Groups:
Their Invariants and Representations}, Princeton University Press,
1939.
%
\bibitem{PopovVinberg}
V.L.~Popov and E.B.~Vinberg,
  \textit{Invariant theory}, in Algebraic
Geometry IV,  Encycl.~Math.~Sci., Vol. 55, Springer-Verlag, 123-273,
1994.
%
%
\bibitem{LindenPopescu}N.~Linden and S.~Popescu,
On multi-particle entanglement, Fortschr.\  Phys.\  {46} (1998),
567--578.
%
\bibitem{GrasslRottelerBeth}M. Grassl, M. R\"{o}tteler, and T. Beth,
Computing local invariants of qubit systems,
Phys.\ Rev.\  A 58 (1998),
1833-1859.
%
\bibitem{KingWelshJarvis}R~C~King, T~A~Welsh and P~D~Jarvis,
The mixed two-qubit system and the structure of its ring of local invariants,
J.\ Phys.\ A: Math.\ Theor.\  {40} (2007), 10083-10108.
%
%
\bibitem{GPKPOMI09}V.~Gerdt, Yu.~Palii and A.~Khvedelidze,
On the ring of local polynomial invariants for a pair of entangled
qubits, J. Math. Sci., 168 (2010), 368-378.
%
%
\bibitem{KusZyczkowski} M.~Kus and K.~\'Zyczkowski,
Geometry of entangled states, Phys.\  Rev.\ A 63, 032307, 13 pages,
2001.
%
\bibitem{BengtssonZyczkowski}
I.~Bengtsson and K.~\'Zyczkowski, {\it Geometry of Quantum States.
An Introduction to Quantum Entanglement}, Cambridge University Press
2006.
%
\bibitem{HioeEberly}F.T.~Hioe and J.H.~Eberly, N-Level Coherence Vector and Higher Conservation
Laws in Quantum Optics and Quantum Mechanics.
Phys.\ Rev.\ Lett.\ 47 (1981), 838-841.
%
%
\bibitem{Djokovic}D.~Djokovic, Poincar\'e series for local unitary invariants of
mixed states of the qubit-qutrit system,  quant-ph/0605018v1 (2006) 5 pages.
%
\bibitem{Jacobson} N. Jacobson, \textit{Lie Algebras}, Wiley-Interscience, New-York, London, 1962.
%
\bibitem{Gelfand} I.M. Gel'fand, The center of the infinitezimal group ring,
Matematicheskii sbornik (English version: Sbornik: Mathematics), Vol. 26, N1 (1950), pp. 103-112.
%
\bibitem{Biedenharn}L.C.~ Biedenharn,
On the representations of the semisimple Lie groups. The explicit
construction of invariants for the unimodular unitary group in N
dimensions. J.\  Math.\ Phys.\ 4 (1963), 436-445.
%
\bibitem{Dalitz}R.H.~Dalitz, Constraints on the statistical tensor
for low-spin particles produced in strong interaction processes,
Nucl.\ Phys.\  87, (1966), 89-99.
%
\bibitem{Minnaert}P.~Minnaert, Spin-density analysis. Positivity
conditions and Eberhard-Good theorem, Phys.\ Rev.\  151, (1966),
1306-1318.
%
\bibitem{DeenKabirKarl}S.M.~Deen, P.K.~Kabir and G.~Karl, Positivity
constraints on density matrices,  Phys.\ Rev.\  D 4, (1971),
1662-1666.
%
\bibitem{Kimura}G~Kimura, The Bloch vector  for N-level systems,
Phys.\ Lett.\ A 314, (2003) 339-349.
%
\bibitem{ByrdKhaneja}M~S~Byrd and N~Khaneja,
Characterization of the positivity of the density matrix in terms of
the coherence vector representation, Phys.\ Rev.\ A 68 (2003)
062322, 13 pages.
%
\bibitem{GKPYadPhys10}V.~Gerdt, A.~Khvedelidze and Yu.~Palii,
Constraints on SU(2)$\otimes$ SU(2) invariant polynomials for a pair
of  entangled qubits, Phys. Atom. Nucl., vol. 74, N6 (2011) 893-900.
%
\bibitem{HochsterRoberts} M. Hochster and J. Roberts, Rings of invariants of reductive groups acting on regular rings are Cohen-Macaulay, Advances in Mathematics 13 (1974) 125-175.

%
\bibitem{Macfarlane} A.J.~Macfarlane and H.~Pfeiffer, On characteristic equations,
trace identities and Casimir operators of simple Lie algebras, J.\ Math.\ Phys.,\ 41 (2000) 3192-3225.
%
\bibitem{OgievetskyPolubarinov}
V.I.~Ogievetsky and I.V.~Polubarinov, Eightfold-way formalism in SU(3) and 10- and 27-plets,
Yad.\ Fiz.\  4 (1965) 853-861.
%
\bibitem{Bukhvostov}
A.P.~Bukhvostov, The algebra of $3\times 3$ matrices in
8-dimensional vector representation, preprint of St-Petersburg
INP-1821/-1822  (1992).


\end{thebibliography}
\end{document}